\documentclass[10pt,aps,prb,twocolumn]{revtex4}
\usepackage{graphicx}
\usepackage{subfigure}
\usepackage{color}

\begin{document}
\title{Nonreciprocal transmission of microwaves through a long Josephson junction}
\author{A. L. Pankratov$^{1,2,3}$, K. G. Fedorov$^{4,5,6}$, M. Salerno$^{7}$, S. V. Shitov$^{5,8}$, and A. V. Ustinov$^{4,5}$}
\affiliation{$^1$Institute for Physics of Microstructures of RAS, Nizhny Novgorod, 603950, Russia}
\affiliation{$^2$Laboratory of Cryogenic Nanoelectronics, Nizhny
Novgorod State Technical University n.a. R.E. Alekseev, 603950 Nizhny Novgorod, Russia}
\affiliation{$^3$Lobachevsky State University of Nizhni Novgorod, 603950 Nizhny Novgorod, Russia}
\affiliation{$^4$Physikalisches Institut, Karlsruhe Institute of Technology, D-76128 Karlsruhe, Germany}
\affiliation{$^5$National University of Science and Technology MISIS, Leninsky prospekt 4, Moscow 119049, Russia}
\affiliation{$^6$Walther-Mei{\ss}er-Institut, Bayerische Akademie der Wissenschaften, D-85748 Garching, Germany}
\affiliation{$^7$Dipartimento di Fisica "E. R. Caianiello", CNISM and INFM Gruppo Collegato di Salerno,
Universit\'{a} di Salerno, Italy}
\affiliation{$^8$Kotelnikov Institute of Radio Engineering and Electronics of RAS, Moscow 125009, Russia}
\begin{abstract}
Nonreciprocal microwave transmission through a long Josephson junction in the flux-flow regime is studied analytically and numerically within the framework of the perturbed sine-Gordon model. We demonstrate that the maximum attenuation of the transmitted power occurs when the direction of the flux flow is opposite to the direction of the microwave propagation. This attenuation is nonreciprocal with respect to the flux-flow direction and can be enhanced by increasing the system length and proper impedance matching of the junction ends to external transmission line.
\end{abstract}

\maketitle

Isolators and circulators transmit microwave power in one direction and do not transmit in the opposite direction. They are used to protect microwave signal sources from harmful effects of standing waves and also guard amplifiers from unwanted signal reflections. These useful functions become possible due to nonreciprocal properties of isolators and circulators, which are gained at price of adding rather bulky microwave components employing magnetic materials such as ferrites.

Recently, a demanding application niche for compact cryogenic microwave isolators has been created by the advances in superconducting qubits and rapid progress in circuit quantum electrodynamics \cite{Clarke-Wilhelm,SG-2008}. Superconducting quantum circuits are operated at millikelvin temperatures and measured using weak microwave signals, which in turn need to be amplified using low-noise cryogenic amplifiers. Bulky conventional isolators and circulators become here rather inconvenient and, moreover, harmful for superconducting circuits due to their relatively large stray magnetic fields. In this area, there is a great need for compact, possibly on-chip, nonreciprocal microwave devices. A possible way towards implementing such a device is based on parametric modulation \cite{Devoret}. An alternative approach to implementing nonreciprocal functionality can employ active transmission line sections with gain in one direction and attenuation in the other. Using a superconducting on-chip flux-flow amplifier \cite{FFT2,Nordman} based on a long Josephson junction (LJJ) is an opportunity along this path. Such amplifiers are expected to have wide freqiency band \cite{Nordman-99} and rather low noise properties \cite{Monaco}.

Our recent experiments have revealed the presence of a notable
non-reciprocity in transmission of a microwave signal through an LJJ biased in the flux-flow regime \cite{FedorovIso_2012}.
This nonreciprocal behavior can be intuitively explained by the interaction of the microwave signal with the moving chain of Josephson fluxons inside the junction. Microwave signal frequency $f_{\rm MW}$ here is typically much lower than the frequency $f_{\rm FF}$ of flux-flow type Josephson oscillations. A preferred non-reciprocal configuration for propagation of the electromagnetic wave is created by choosing a specific direction of the flux flow given by a combination of polarities of the applied bias current and the in-plane magnetic field. The microwave transmission from one end of LJJ to the other is enhanced when the wave vector of the applied microwave coincides with the direction of the flux-flow. In contrast to this, the propagation is damped when the microwave signal is applied to the fluxon's exit port of LJJ. Thus, LJJ acts as an on-chip isolator for external microwave signals, with its transmission properties being fully controlled by the bias current and in-plane magnetic field. The discussed isolation functionality is somewhat close to the working principle of traveling wave isolators proposed for optical applications \cite{OptoAcoIso-1,OptoAcoIso-2}. Previous experiments have shown \cite{Koshelets-APL-1996,Koshelets-SST-2000} that the emission of Josephson radiation from LJJ in the
flux-flow regime (at the frequency of flux flow $f_{\rm FF}$) is negligible from the side where the fluxons enter LJJ. For this reason, the unwanted emission of such an LJJ isolator back into the incoming microwave line can be neglected.

The above described operation principle offers an opportunity to
create an isolator for microwave cryogenic applications. However, to fulfill the task of constructing a practically useful isolator, one needs
to achieve the isolation level comparable to the standard $20-25$
dB or better. This benchmark remained beyond reach in the first experiment \cite{FedorovIso_2012}. To achieve the isolation level needed for useful applications, it is required to better understand the physics of up- and down-conversion processes of microwaves inside LJJ. As the typical impedance of LJJ is much lower than that of the external circuits, one also has to study the matching conditions of LJJs to an external network.

This paper presents a systematic numerical and analytical study of nonreciprocal microwave transmission through LJJ in the flux-flow regime. We study nonreciprocal microwave transmission through a LJJ depending on the system parameters and impedance matching conditions.

A qualitative understanding of the nonreciprocal effect can be gained from the analysis of the following  perturbed sine-Gordon equation (PSGE)
\begin{equation}
\label{SGE}
{\phi}_{tt}+\alpha{\phi}_{t}
-{\phi}_{xx}=\beta{\phi}_{xxt}+\eta-\sin (\phi),
\end{equation}
with the boundary conditions
\begin{equation}
\phi_{x}(0,t)= \Gamma + A \sin\Omega t, \;\; \phi_{x}(L,t)= \Gamma,
\label{boundary}
\end{equation}
as a model of a LJJ operating in the flux-flow regime.
Here indices $t$ and $x$ denote temporal and spatial derivatives,
$\phi$ is the Josephson phase difference. Space and time are normalized
to the Josephson penetration length $\lambda _{J}$ and to the
inverse plasma frequency $\omega_{p}^{-1}$, respectively,
$\alpha={\omega_{p}}/{\omega_{c}}$ is the damping parameter,
$\omega_p=\sqrt{2eI_c/\hbar C}$, $\omega_{c}=2eI_cR_{N}/\hbar$,
$I_c$ is the critical current, $C$ is the LJJ capacitance, $R_N$ is
the normal state resistance, $L$ is the dimensionless length of the junction in units of $\lambda _{J}$, $\beta$ is the surface loss parameter,
$\eta$ is the dc bias current density, normalized to the
critical current density $J_c$ (here $e$ is the electron
charge, $\hbar$ is the Planck constant).

In the boundary conditions given by Eq. (\ref{boundary}), $\Gamma$ denotes dc in-plane magnetic field at the edges of the junction normalized to $\lambda _{J}J_{c}$, while the ac term with normalized amplitude $A$ and frequency $\Omega=2\pi f_{\rm FF}/\omega_p$  applied at the $x=0$ boundary accounts for a microwave radiation applied to the junction. It is well known that the flux-flow regime is achieved when $\Gamma>2$ and is characterized by an average of  $ N=\Gamma L/2\pi$ fluxons moving in the direction fixed by the signs of $\Gamma$ and $\eta$ on a uniform rotating background $\phi_{0}= \omega t + \Gamma x$. The nonreciprocal effect must be related  to a different dynamical behavior of the radiation generated at the $x=0$ boundary when traveling inside the junction in the same flux-flow direction or against to it, this depending on the sign of $\Gamma$. In this respect, it is convenient to separate the flux-flow background $\phi_0$ from the rest of the field and  derive an effective field equation fulfilling the reflective boundary conditions. Note that, the linear increase in space of the background $\phi_0$ allows satisfying the dc part of the boundary conditions (\ref{boundary}).  In this approximation one also reproduces the resistive branch  $\eta = \alpha \omega$ of the $I-V$ characteristic of the Josephson junction. To account for the ac part of the boundary condition we assume the following ansatz solution
\begin{equation}
\phi(x,t)=\phi_{0}+f_{+}(x)\cos\Omega t+f_{-}(x)\sin\Omega t+\psi(x,t)+\theta
\label{Phi_ansatz}
\end{equation}
where $f_\pm$ are space dependent functions to be fixed so to satisfy the ac part of the boundary condition (\ref{boundary}), $\theta$ is an arbitrary initial phase and  $\psi(x,t)$ represents the radiation field inside the junction.
\begin{figure}[ptbh]
\includegraphics[width=0.5\textwidth]{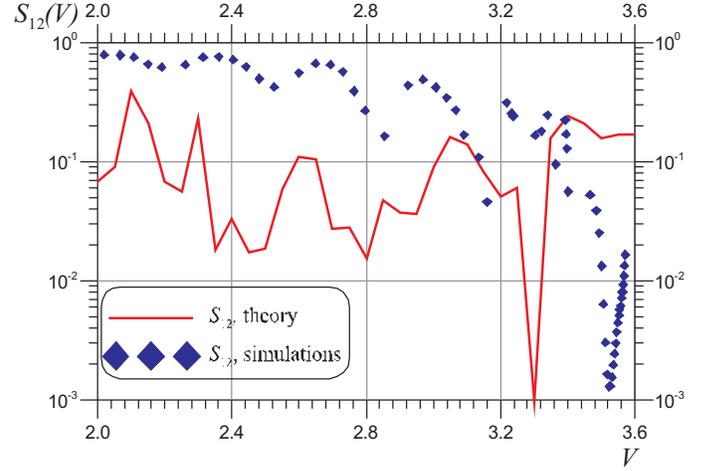}
\caption{Transmitted power $S_{12}$ for $\alpha=0.1$, $L=20$, $\Gamma=4$, $\omega=0.231$ and $A=0.5$.}
\label{FigTheo}
\end{figure}
To simplify the analysis we neglect the surfaces losses in the PSGE, so we put $\beta=0$ in Eq. (\ref{SGE}). It can be readily checked that by substituting Eq. (\ref{Phi_ansatz}) into Eq. (\ref{SGE}) one obtains
\begin{eqnarray}\label{SGEpsi}
\psi_{xx} &-& \psi_{tt} - \alpha \psi_{t}= \alpha \omega - \eta  +  \\ &&
\sin \left[\phi_0 + f_{+}\cos \Omega t+ f_{-}\sin \Omega t + \psi + \theta\right] \nonumber
\end{eqnarray}
with $\psi$ satisfying the reflective boundary conditions
\begin{equation}
\psi_{x}(0,t)= \psi_{x}(L,t) = 0,
\label{linearbc}
\end{equation}
providing the functions $f_\pm (x)$ are fixed as
\begin{equation}
f_{\pm}(x)=\mp \frac{A}{\Omega \sqrt{\alpha^2+\Omega^2}}
\frac{h^{\pm}(x_+,x_-)+h^{\pm}(x_-,x_+)}{\cos(2 L \Omega^{+})-\cosh(2 L\Omega^{-})}
\label{Apm}
\end{equation}
with $\Omega_{\pm}=\left[\frac{\Omega}{2}(\sqrt{\alpha^2+\Omega^2} \pm \Omega) \right]^{\frac 12}$,
$\; x_\pm= L \pm (x-L)$ and
\begin{equation}
h^{\pm}(x,y)=\Omega_{\pm}\cos\Omega_{+}y\sinh\Omega_{-}x\pm\Omega_{\mp}\cosh\Omega_{-}x\sin\Omega_{+}y. \nonumber
\end{equation}
Assuming the field $\psi$ small,  Eq. (\ref{SGEpsi}) can be linearized as
\begin{eqnarray}
\psi_{xx} - \psi_{tt} - \alpha \psi_{t}= \alpha \omega - \eta  + \nonumber \\  \sum_{m=-\infty}^{+\infty} J_m (f(x)) \sin(\tilde \omega_m + \Gamma x + m \Phi(x) + \theta) + \nonumber \\
\sum_{m=-\infty}^{+\infty} \psi J_m (f(x)) \cos(\tilde \omega_m + \Gamma x + m \Phi(x) + \theta).
\label{linearizedSGE}
\end{eqnarray}
where we have introduced the functions
\begin{equation}
f(x)=\sqrt{f_{+}(x)^2 + f_{-}(x)^2},\;\;  \Phi(x)=\tan^{-1}{\frac{f_{+}(x)}{f_{-}(x)}},
\label{Aphi}
\end{equation}
and denoted with $J_m(\beta)$ the Bessel function of the order $m$ and  $\tilde \omega_m= \omega + m \Omega$. A solution of Eq. (\ref{linearizedSGE}) satisfying the boundary condition (\ref{linearbc}) can be obtained as a Fourier series in the form (see also \cite{SS00})
\begin{equation}
\psi(x,t) = \sum_n\sum_m \left[B^{+}_{nm} c^{+}_{m}(t)+B^{-}_{nm} c^{-}_{m}(t)\right]\cos k_n x,
\label{radfiels}
\end{equation}
\begin{figure}[ptbh]
\includegraphics[width=0.5\textwidth]{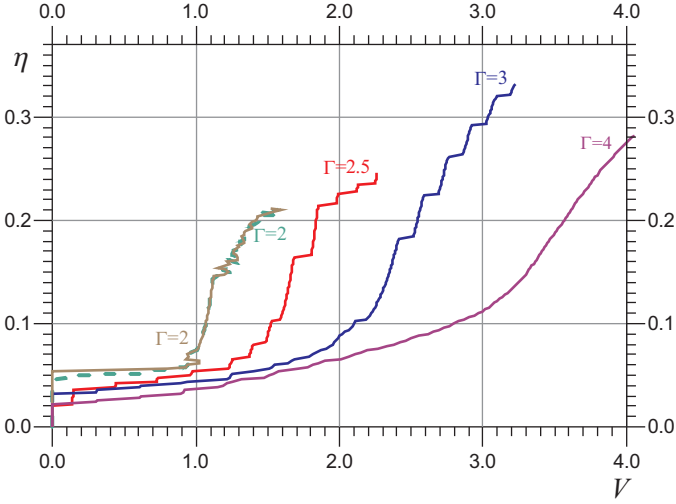}
\caption{Current-voltage characteristic of a LJJ for the mismatched case, $r_{L}=r_{R}=20$, $c_{L}=c_{R}=10$, and $L=20$.}
\label{CV}
\end{figure}
with coefficients  $B^{\pm}_{nm}$ of the expansion  given by
\begin{eqnarray}
&&
B^{\pm}_{nm}=\frac{(\omega_m^2-k_n^2)I^{\mp}_{nm}\pm\alpha\omega_m I^{\pm}_{nm}}{(\omega_m^2-k_n^2)^2+ \alpha^2 \omega_m^2 } \nonumber \\ &&
I^{\pm}_{nm}=\frac 1L\int_0^L\cos(k_n x) J_m(f(x)) s^{\pm}_m(x) dx,
\end{eqnarray}
where  $c^{+}_{m}(t)=\cos(\omega_m t + \theta)$, $c^{-}_{m}(t)=\sin(\omega_m t + \theta)$,   $s^{+}_m(x)=\cos(\Gamma x + m \Phi(x))$, $s^{-}_m(x)=\sin(\Gamma x + m \Phi(x))$ and we  denoted  $k_n= n \pi/L$.

From Eq. (\ref{radfiels}) one can obtain the amplitudes of the oscillations at the edges of the junction to estimate the gain. While this expression  is complicated to manipulate analytically, it can  be  evaluated numerically by truncating the series. Let us consider $S$ coefficients, where $S_{12}$ corresponds to the direct (supplied from the left end) signal and $S_{21}$ corresponds to the reversed (supplied from the right end) signal, normalized to the amplitude of the input driving $A$. In Fig. \ref{FigTheo} we compare the transmitted power $S_{12}$ computed from the Fourier modes of the radiation signal (\ref{radfiels}) with the one obtained from the numerical simulations. Although the theory slightly overestimates the phenomenon and the main minimum is shifted to a smaller value of $V$, there is a clear evidence of the existence of the nonreciprocal effect with a good qualitative agreement between analytical and numerical results.

To associate our results with the experimental data it is interesting to consider a more realistic situation in which the surface losses and boundary loads are included into the model. In this case, however, an analytical treatment is out of reach and we shall recourse to the direct numerical simulations of the PSGE with boundary conditions
\begin{eqnarray}\label{x=0}
\phi(0,t)_{x}+r_L c_L\phi(0,t)_{xt}-c_L\phi(0,t)_{t t}+\\
\beta r_L c_L\phi(0,t)_{xtt}+\beta\phi(0,t)_{x
t}=\Gamma-\Delta\Gamma+\Gamma_{12}(t), \nonumber \\ \phi(L,t)_{x}+r_R
c_R\phi(L,t)_{x t}+c_R\phi(L,t)_{t t}+ \label{x=L} \\  \beta r_R
c_R\phi(L,t)_{xtt}+\beta\phi(L,t)_{x t}=\Gamma+\Delta\Gamma+\Gamma_{21}(t),
\nonumber
\end{eqnarray}
appropriate for $RC$ loads \cite{Parment,pskm}.
Here $\Delta\Gamma$ is a small magnetic field difference, $\Gamma_{12}(t)=A\sin(\Omega t)$ and $\Gamma_{21}(t)=A\sin(\Omega t)$ are ac magnetic fields which are supplied either from the left ($\Gamma_{12}$) or from the right ($\Gamma_{21}$) junction ends, respectively, but not from both ends simultaneously.
\begin{figure}[ptbh]
\subfigure[\label{FigSLGun}]
{\includegraphics[width=0.5\textwidth]{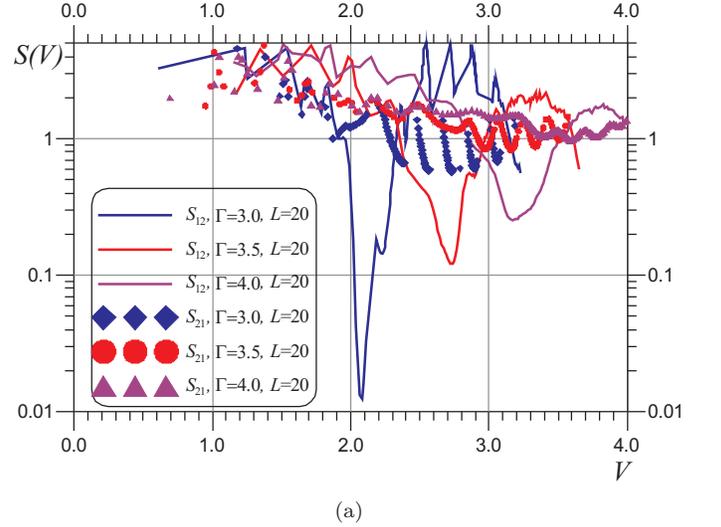}}
\subfigure[\label{FigSLGL80}]
{\includegraphics[width=0.5\textwidth]{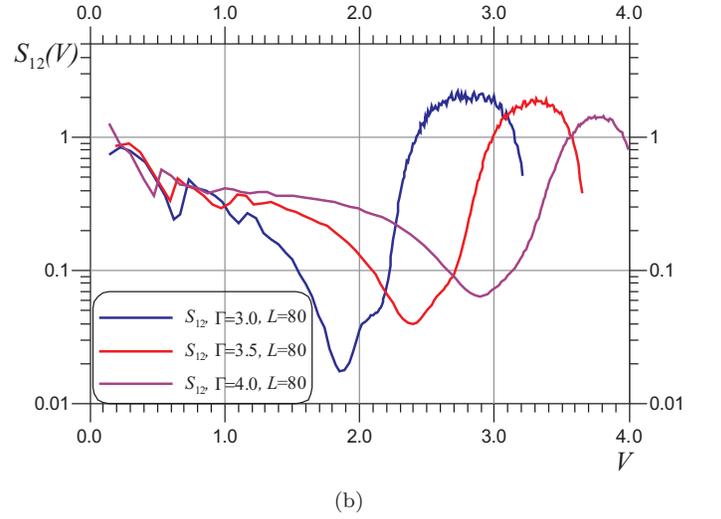}}
\caption{(a) The transmitted power $S_{12}$ and $S_{21}$ for $\alpha=0.03$ and
$\Gamma=3.0$, $\Gamma=3.4$, $\Gamma=4.0$; $r_{L}=r_{R}=20$,
$c_{L}=c_{R}=10$. (b) The transmitted power $S_{12}$ for $\alpha=0.03$ and
$\Gamma=3.0$, $\Gamma=3.5$, $\Gamma=4.0$; $r_{L}=r_{R}=20$,
$c_{L}=c_{R}=10$, $L=80$.}
\end{figure}
The dimensionless resistances and capacitances, $r_{L,R}$ (normalized on the characteristic impedance of the junction $Z_0$) and $c_{L,R}$ (normalized to the capacitance $C_0=1/\omega_p Z_0$), are the LJJ RC-loads placed at the left (output) and at the right (input) ends, respectively \cite{Parment}.

We are going to investigate the nonreciprocal microwave transmission by the direct simulation of Eq. (\ref{SGE}) with the boundary conditions (\ref{x=0}). Results are compared with both our simplified analytical study (in the corresponding range of validity, see Fig. \ref{FigTheo}) and with the experiment performed using a device fabricated in a standard Nb-AlO$_x$-Nb technology \cite{FedorovIso_2012}.

The settings used in our numerical study are the following. The
junction length vary from $L=20$ to $L=80$, the damping is
$\alpha=0.03$, the surface losses are $\beta=0.03$. The sign of the bias current and the
magnetic field are chosen such that the fluxons are moving from right to
left and the radiation is emitted from the left end of the junction.
To supply the maximum ac power to the LJJ, it should be well-matched to the external transmission line, so the values of $r_L$ and $r_R$ must vary from 0.5 to 2.
\begin{figure}[ptbh]
\subfigure[\label{FigSacAL20}]
{\includegraphics[width=0.5\textwidth]{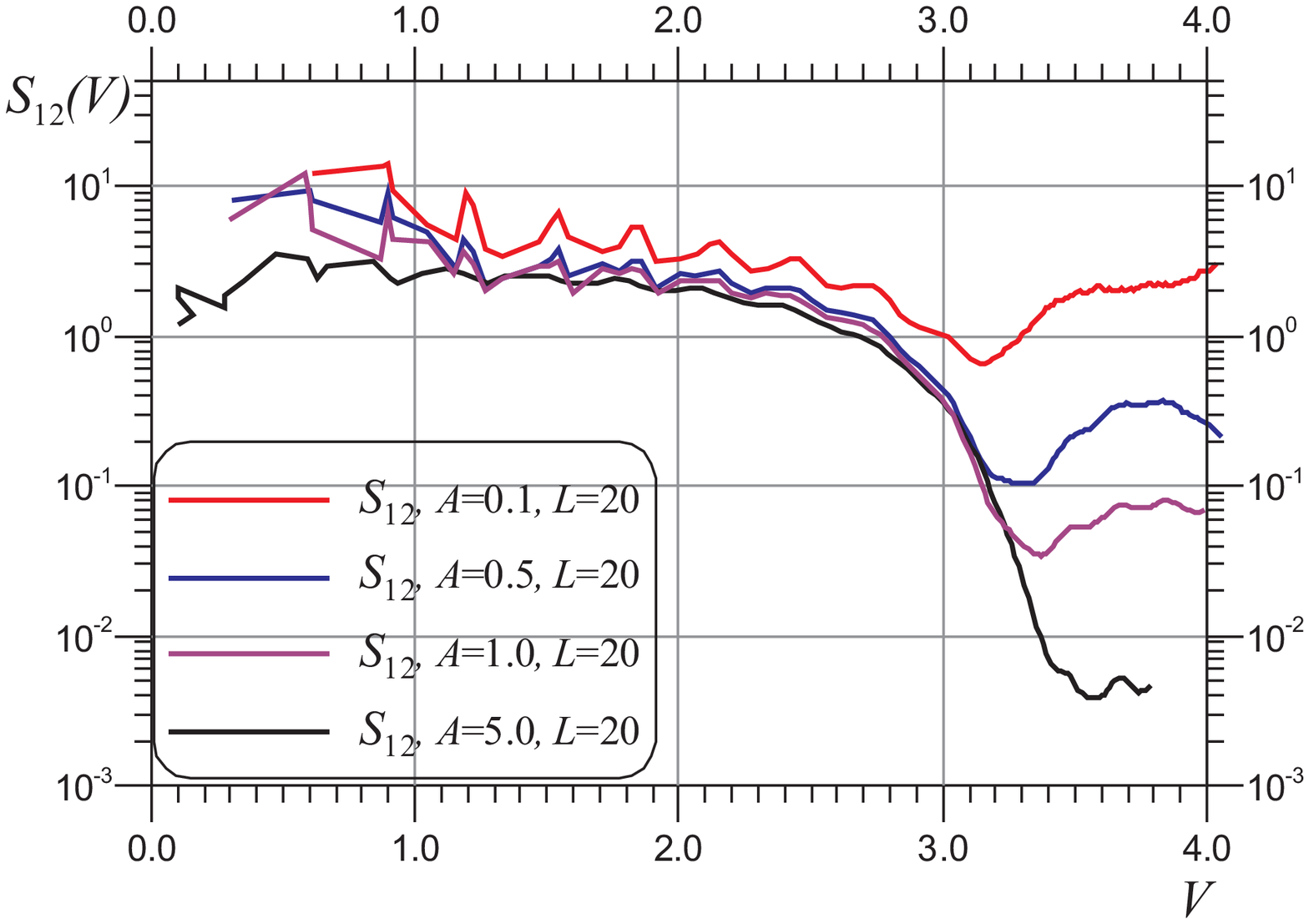}}
\subfigure[\label{FigSacRC}]
{\includegraphics[width=0.5\textwidth]{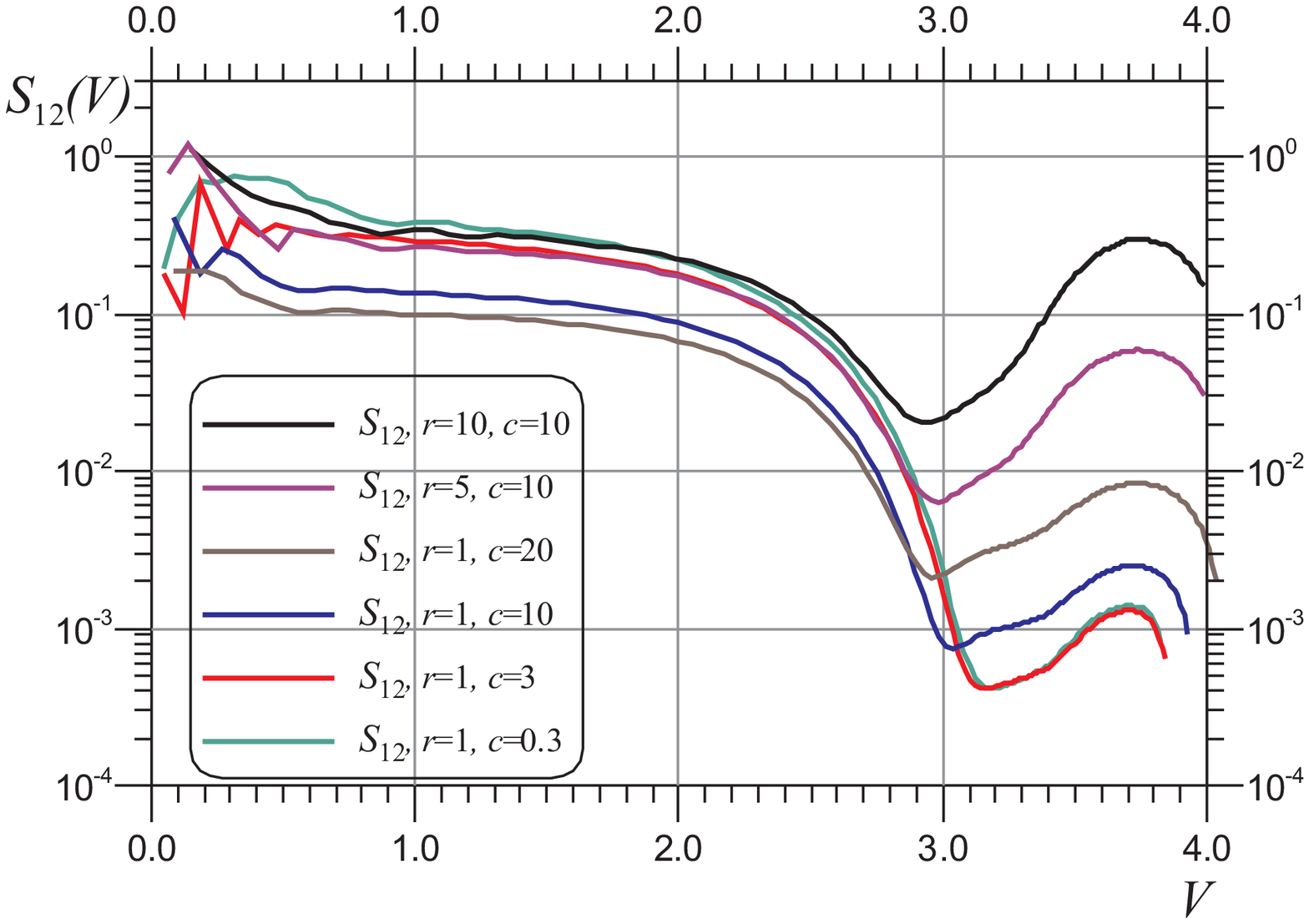}}
\caption{(a) The transmitted power $S_{12}$ for $\alpha=0.03$, $L=20$,
$\Gamma=4$, $r_L=r_R=10$ and $c_L=c_R=10$ versus driving amplitude.
(b) The transmitted power $S_{12}$ for $\alpha=0.03$, $L=80$, $A=0.5$ and
$\Gamma=4$ versus load resistance $r=r_L=r_R$ and capacitance $c=c_L=c_R$.}
\end{figure}
In reality, it is difficult to achieve a proper matching due to technical limitations of fabrication processes, so to investigate the physics of the considered effect we will vary $r_L$ and $r_R$ in a broad range, but let us start from the poorly matched case $r_L$=$r_R$=20 and the short junction length $L=20$.
The ac signal frequency is taken fixed at $\Omega=0.1$ close to the
experimental value, while its amplitude is set at $A=0.5$ in the most cases except
Fig. \ref{FigSacAL20}, where dependence on the driving amplitude is studied.

The current-voltage characteristics for $L=20$, $r_{L}=r_{R}=20$, $c_{L}=c_{R}=10$
and different values of magnetic field $\Gamma$ (see Fig.
\ref{CV}), look similar to usual experimental curves: one can see the
displaced linear slope at $\Gamma=2.0$, the Fiske steps at $\Gamma=2.5$
and $\Gamma=3.0$, which are smoothed due to surface losses at larger
magnetic fields $\Gamma=4.0$ (the flux-flow steps). All these curves are calculated
for the case where ac driving is supplied at the left (output) LJJ
end, while the dashed curve is for $\Gamma=2.0$, and the ac
driving supplied from the right (input) junction end. Curves for the
higher magnetic fields and the ac driving from the right end are not shown
since they nearly coincide with the shown $I-V$ characteristics for the same magnetic
fields, because the weak ac signal does not affect the $I-V$ characteristics.

Analyzing coefficients of microwave power transmission through the junction $S$ coefficients, one can see from Fig. \ref{FigSLGun} that in the case of small magnetic field $\Gamma$ (soft vortex chain), the nonreciprocal effect is rather large, and the $S_{12}$-parameter can be around two orders of magnitude.
\begin{figure}[ptbh]
{\includegraphics[width=0.5\textwidth]{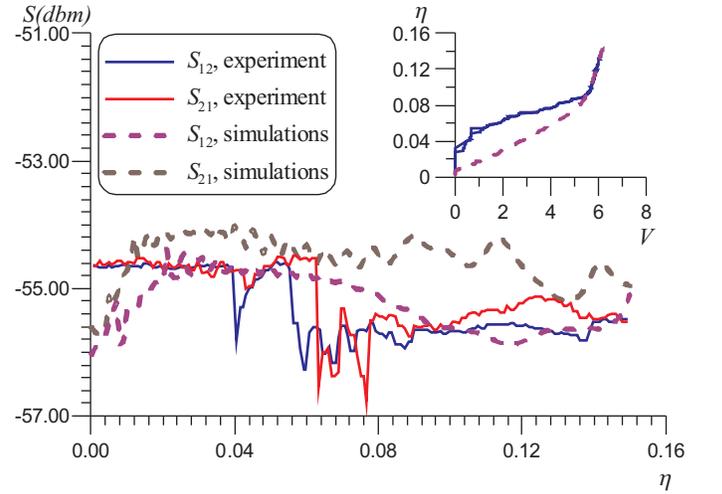}}
\caption{The experimentally measured transmitted power $S_{12}$ and $S_{21}$ (solid curves) compared to the numerically simulated data (dashed curves)
for $L=18$, $\alpha=0.014$, $\Gamma=7.3$, $r_L=r_R=10$, $c_L=c_R=10$, $\beta=0.007$, $\Delta \Gamma=0.2\Gamma$, $\omega=0.231$ and $A=0.5$. Inset:
comparison of the current-voltage curves for the same parameters. }
\label{FigExp}
\end{figure}
For large magnetic field $\Gamma=4$ the $S_{12}$-parameter decreases roughly by 50$\%$ and becomes comparable to the measured experimental values \cite{FedorovIso_2012} reported for similar range of parameters. From our numerical studies we also obtain clear indications on how the nonreciprocal  effect could be enhanced by properly choosing the parameters. In this respect, we have first investigated the dependence on the junction length $L$, keeping all other parameters the same. Considering the behavior of $S_{21}$-parameter (symbols in Fig.
\ref{FigSLGun}) one can see that at large magnetic fields the curves slightly oscillate around unity. Note, that the $I-V$ curves for $L=80$ for various values of magnetic field $\Gamma=2.0$, $\Gamma=2.5$, $\Gamma=3.0$ reach larger values of current than in Fig. \ref{CV} for $L=20$.
From Fig. \ref{FigSLGL80} one can see that the ac signal attenuation
for larger lengths in the area of large magnetic field $\Gamma$ becomes stronger,
while for smaller $\Gamma$ the area of minimal $S_{12}$ increases.
For the particular value of the in-plane magnetic field $\Gamma=4$, we consider the dependence of the effect on the ac signal amplitude, see Fig. \ref{FigSacAL20}. One can observe the paradoxical behavior of $S_{12}$: the nonreciprocal effect becomes stronger with increase of the driving amplitude. However, the analysis of the power spectral density for various values of the amplitude clarifies the situation. With increase of the ac input driving amplitude, the signal amplitude at the opposite end at the basic frequency $\Omega$ becomes almost constant up to $A=5$, however due to the nonlinearity of the Josephson junction the amplitudes of higher harmonics (at $\Omega$, 2$\Omega$, 3$\Omega$) increase. When calculating $S_{12}$, only the amplitude at frequency $\Omega$ is taken into account, which leads to $S_{12}$ decrease. It should be noted that for the considered parameters and the amplitudes above $A=5$, the nonreciprocal effect decreases (the minimum value of $S_{12}$ grows). The nonreciprocal effect can be further improved by better impedance matching of the LJJ with an external waveguide system. In Fig. \ref{FigSacRC}
several curves of $S_{12}$ for the various values of load resistance
$r=r_L=r_R$ and capacitance $c=c_L=c_R$ are presented. Here the power of the propagated ac signal can be suppressed almost up to three orders of magnitude even for large magnetic fields $\Gamma=4$, so the isolation can be greatly improved. As one can see,
the chosen value of the load capacitance $c_L=c_R=10$ is not the
optimal one, since stronger attenuation of the transmitted signal is
observed at smaller values of the load capacitance. This means that the matching of the structure must be performed in the range of the transmitted signal frequency, while the good matching in the range of flux-flow generation of LJJ is not required.

Figure \ref{FigExp} compares the numerical data (dashed curves) with
the experimentally measured transmitted power $S_{12}$, $S_{21}$ taken
from Ref. \onlinecite{FedorovIso_2012} (solid curves). There is a qualitative agreement between the $S$-parameters.
From our study we expect that by increasing the junction length $L$ and by improving the matching at the frequencies of the propagating signal, one should be able to improve the ac signal isolation up to 20-30 dB in future experiments.

In conclusion, we have investigated, both analytically and numerically, the nonreciprocal microwave transmission through a long Josephson junction in the flux-flow regime within the framework of the sine-Gordon equation. It is demonstrated that the maximum attenuation of the transmitted power occurs when the direction of the fluxon motion is opposite to the direction of a microwave propagation. This attenuation is nonreciprocal in respect to the flux-flow direction and can be enhanced for larger lengths of the junction and by its proper impedance matching to the external microwave network.

This work was supported by the Ministry for Education and Science of Russian Federation under contract no. 11.G34.31.0062 and 3.2054.2014/K, and in the framework of Increase Competitiveness Program of NUST (MISiS) under contract no. K2-2014-025 and of Lobachevsky NNSU under contract no. 02.B.49.21.0003. MS acknowledges partial support from MIUR through a PRIN-2010 initiative.

\end{document}